\documentclass[aps,prd,twocolumn,floatfix,showpacs,eqsecnum]{revtex4-1}

\pdfoutput=1
\usepackage[centertags]{amsmath}
\usepackage{amssymb}
\usepackage{latexsym}
\usepackage{enumerate}
\usepackage{graphicx}
\usepackage{mathrsfs}
\usepackage{hyperref}
\usepackage{stmaryrd}

\usepackage{soul}

\newcommand{\bi}{\begin{itemize}}
\newcommand{\ei}{\end{itemize}}

\newcommand{\be}{\begin{equation}}
\newcommand{\ee}{\end{equation}}

\newcommand{\ba}{\begin{eqnarray}}
\newcommand{\ea}{\end{eqnarray}}

\newcommand{\nn}{\nonumber}

\newcommand{\apjl}{ApJL}

\newcommand{\aj}{AJ}
\newcommand{\mnras}{MNRAS}

\newcommand{\aap}{A\&A}

\newcommand{\nature}{Nature}

\begin{document}

\title{Tidal disruption of a star in the Schwarzschild spacetime: \\
relativistic effects in the return rate of debris} 

\author{Roseanne M. Cheng}
\email{rcheng@physics.gatech.edu}
\affiliation{Center for Relativistic Astrophysics, School of Physics, Georgia Institute of Technology, Atlanta, GA 30332-0430, USA;\\ Visitor at the Kavli Institute for Theoretical Physics, \\ University of California, Santa Barbara, CA 93106-4030, USA}

\author{Tamara Bogdanovi\'c}
\email{tamarab@gatech.edu}
\email{Alfred P. Sloan Research Fellow}
\affiliation{Center for Relativistic Astrophysics, School of Physics, Georgia Institute of Technology, Atlanta, GA 30332-0430, USA;\\ Visitor at the Kavli Institute for Theoretical Physics, \\ University of California, Santa Barbara, CA 93106-4030, USA}

\begin{abstract}

Motivated by an improved multi-wavelength observational coverage of
the transient sky, we investigate the importance of relativistic
effects in disruptions of stars by non-spinning black holes (BHs).
This paper focuses on calculating the ballistic rate of return of
debris to the black hole as this rate is commonly assumed to be
proportional to the light curve of the event.  We simulate the
disruption of a low mass main sequence star by BHs of varying masses
($10^5,10^6,10^7 M_\odot$) and of a white dwarf by a $10^5 M_\odot$
BH. Based on the orbital energy as well as angular momentum of the
debris, we infer the orbital distribution and estimate the return rate
of the debris following the disruption. We find two signatures of
relativistic disruptions: a gradual rise as well as a delayed peak in
the return rate curves relative to their Newtonian analogs. Assuming
that the return rates are proportional to the light curves, we find
that relativistic effects are in principle measurable given the
cadence and sensitivity of the current transient sky surveys.
Accordingly, using a simple model of a relativistic encounter with a
Newtonian parametric fit of the peak leads to an overestimate in the
BH mass by a factor of $\sim {\rm few}\times0.1$ and $\sim {\rm few}$
in the case of the main sequence star and white dwarf tidal
disruptions, respectively.

\end{abstract}
 

\maketitle

\section{Introduction}
\label{sec:intro}

In a tidal disruption event, a star on a marginally bound orbit
disrupts as it reaches a limiting distance of closest approach known
as the tidal radius.  After disruption one fraction of stellar debris
is launched towards the BH and remains bound to it, while the
remaining fraction is launched away from the BH and unbound.  Bound
debris forms an accretion disk around the BH and generates the key
signature of these events: the tidal disruption flare.  Early
theoretical models of tidal disruptions predict that the accretion rate
onto the BH and consequently the tidal disruption flare, decays with a
characteristic time dependence $\propto t^{-5/3}$
\cite{rees1988,phin1989,evan1989}. This prediction led to the first
detections of tidal disruption candidates in the UV and soft X-ray
band
\cite{bade1996,grup1999,komo1999,grei2000,donl2002,komo2004,halp2004}
and has been used as a diagnostic for the presence of massive BHs in
previously inactive galaxies ever since.

The success of the early theory of tidal disruptions is remarkable,
especially given that predictions of the light curve properties are
based on the Keplerian return rate of the debris and do not account
for the complexities of accretion and radiative processes.
Furthermore, subsequent studies have found deviations from the
$t^{-5/3}$ fall-off for different stellar structure of disrupted stars
\cite{loda2009,guil2013}. It has been shown for partially disrupted
stars (i.e., close encounters with the BH just outside the tidal
radius) that the presence of a self-gravitating remnant core modifies
the dynamics of the debris and hence, the accretion rate onto the BH
\cite{guil2013}. In this scenario, the released debris elements are
deflected from orbits that they would otherwise have in the absence of
self-gravitating remnant, resulting in a slope different from $t^{-5/3}$. Similarly, a different dependence is expected for stars
disrupted on bound (as opposed to marginally bound, parabolic) orbits
\cite{haya2013,dai2013}.

An additional layer of complexity in linking the calculated return
rate of debris to the observed light curves stems from the fact that
the majority of tidal disruption events are expected to result in
super-Eddington accretion rates. If these high accretion rates give
rise to super-Eddington luminosities, then radiative feedback is also
expected to affect the appearance of the tidal disruption light curves
\cite{loeb1997,ulme1999,strub2009,strub2011,loda2010}.  Finally,
disruptive encounters that occur very close to BHs are subject to
relativistic effects, leaving an intriguing possibility that
information about the spacetime of a BH is imprinted in the light
curve \cite{kesd2012,haas2012,east2013}.

Given the large number of physical processes that play a role in tidal
disruption events it was somewhat fortunate that the early
observational campaigns, which relied on a relatively sparse sampling
of the light curves, were immune to this level of complexity.
Improved multi-wavelength observational coverage of the transient sky
with GALEX, Swift, Pan-STARRS1, PTF, CRTS, ASAS-SN in the present, and
eROSITA, LSST in the future (and the proposed mission ISS-Lobster),
are opening a window to study tidal disruption events in unprecedented
detail
\cite{vanv2011,bloo2011,leva2011,burr2011,cenk2012,geza2012,camp2013,kann2013,chor2014,arca2014,holo2014}.
For the first time high cadence observations in
principle permit the measurement of relativistic effects in the light
curves of tidal disruptions. This possibility motivates our work on
stellar disruptions in the relativistic regime.

In this paper, we investigate the relativistic effects in the return
rate of the debris for disruptive encounters between main sequence
(MS) stars and BHs of mass $10^5, 10^6, 10^7 M_\odot$. MS-BH
encounters are expected to dominate the detection rates and given a
large sample of tidal disruption event candidates in the near future,
the likelihood of observing relativistic encounters is also higher for
this class of events.  We also investigate partially disruptive,
relativistic encounters between a white dwarf (WD) and a $10^5
M_\odot$ intermediate mass black hole (IMBH) with periastrons very
close to the BH.  They are important because if detected, they will
uncover the otherwise hidden IMBH population
\cite{ross2009,shch2013,macl2014}.

In this study, we use a hydrodynamics code coupled with self-gravity
as well as a general relativistic treatment of the tidal interaction
to study the disruption process in the frame of the star
\cite{chen2013}.  We also develop a series transformation to map
quantities in the local frame to the black hole frame.  This allows us
to characterize the evolution in the distribution of orbital energy
and angular momentum of the disrupted star and obtain the ballistic
debris return rate onto the BH.  We find that relativistic effects are
significant and in principle measurable, assuming that the ballistic
return rate is proportional to the light curve.

The remainder of this paper is organized as follows: in
Sec.~\ref{sec:params} we introduce the parameters of the model and in
Sec.~\ref{sec:formalism} present the method for mapping the debris in
the frame of the star and the BH frame.  We present the results of the
simulations in Sec.~\ref{sec:results}, discuss their implications in
Sec.~\ref{sec:discussion} and conclude in Sec.~\ref{sec:conclusions}.

\section{Parameters of the model}
\label{sec:params}

Consider a star of mass $M_*$ and radius $R_*$ in an orbit about a BH
of mass $M$ and define the mass ratio as $\mu \equiv M_*/M$.  We
characterize the strength of the tidal encounter with a parameter
$\eta$ which compares the hydrodynamical stellar timescale $t_* =
(R_*^3/GM_*)^{1/2}$ to the orbital timescale $t_p = (R_p^3/GM)^{1/2}$
at periastron $R_p$,
\be \label{eqn:encounter_strength}
\eta \equiv t_p/t_* = \left (\frac{R_p^3}{GM}
\frac{GM_*}{R_*^3} \right )^{1/2}
\ee
\cite{pres1977}.  Disruption occurs for $\eta=1$ at the tidal radius
$R_T\equiv R_* \mu^{-1/3}$.  The ratio of stellar radius to periastron
in terms of these parameters is $R_*/R_p = \mu^{1/3}\eta^{-2/3}$.
Each encounter can also be characterized by a penetration factor
$\beta=R_T/R_p = \eta^{-2/3}$.  For $\beta>1$, periastron is less than
the tidal radius and the orbital timescale $t_p$ is smaller than the
hydrodynamical stellar timescale $t_*$.

The variation in the specific orbital energy of the released gas
depends on the change in the Newtonian BH potential $\Phi(r)=-GM/r$
across the diameter of the star at the tidal radius
\cite{rees1988,ston2013}.  For a fluid element with orbital velocity
$\vec{V}$, the specific orbital energy is $\epsilon_N = \frac{1}{2}V^2 + \Phi(r)$.  
The spread in $\epsilon_N$ across a stationary star in
the gravitational potential is $\Delta\epsilon_N = \Phi(R_T-R_*) -
\Phi(R_T+R_*)$.  Expanding this difference to first order one gets
the Newtonian spread in specific energy across the stellar radius at
the tidal radius,
\be\label{eqn:Newtonian_spread_of_energy}
\Delta\epsilon_N = \frac{GM}{R_T} \frac{R_*}{R_T}. 
\ee
For full disruptions, the kinetic energy of the expanding
debris is much larger than the adiabatically decreasing internal
energy and diminishing self-gravitational energy.  By neglecting the
self-gravity of the star in this limit, it can be shown that the debris most
tightly bound to the BH has orbital energy $\Delta\epsilon_N$.  This
approximation is valid for deeply penetrating encounters $\beta\gg 1$
($\eta\ll 1$), where self-gravity is negligible compared to the
specific orbital energy at periastron \cite{ston2013}.  Thus, for full
disruptions, it is reasonable to approximate a ``flat'' distribution
in mass and energy of $dM/d\epsilon \simeq M_*/2\Delta\epsilon_N$ for
the debris \cite{evan1989}.  Assuming that the debris is locked into
Keplerian trajectories after disruption, the semi-major axis of the
most tightly bound debris is $a_m = GM/(2\Delta\epsilon_N) =
R_p^2/(2R_*)$ with a minimum Keplerian return period of $P_m =
(\pi/\sqrt{2GM}) R_p^3 R_*^{-3/2}$.  The rate at which each debris
element returns to its individual periastron after one orbit is
\be\label{eqn:Newtonian_estimated_accretion_rate} 
\dot{M} = \frac{dM}{d\epsilon}\frac{d\epsilon}{dt}  = \frac{1}{3} \frac{M_*}{P_m}\left (\frac{t}{P_m} \right )^{-5/3},
\ee
where the rate of change of specific orbital energy is $d\epsilon/dt =
(1/3)(2\pi GM)^{2/3} t^{-5/3}$.  Equation
(\ref{eqn:Newtonian_estimated_accretion_rate}) shows the canonical form
of the rate at which bound debris returns to the BH after one
orbit \cite{rees1988,phin1989,evan1989}.  The
luminosity $L$ of the flare associated with the accretion disk that
forms from the debris is commonly assumed to be proportional to
$\dot{M}$ as $L \propto \dot{M} c^2$ (although, see
Sec.~\ref{sec:intro} for discussion of difficulty in linking the
debris return rate to the observed luminosity).

\section{Numerical Method}
\label{sec:formalism}

The disruption is simulated in a local moving frame known as Fermi
normal coordinates (FNC), which follows the star along a parabolic
orbit in the known spacetime of a Schwarzschild BH \cite{chen2013}.
The central feature of this method is the use of a relativistic tidal
field in the vicinity of the trajectory.  In the FNC coordinates the
tidal field is given by an expansion of the BH spacetime that
includes quadrupole and higher multipole moments and relativistic
corrections.  This approach allows a Newtonian treatment of the
self-gravity and hydrodynamics of the star while accounting for
relativistic effects inherent to the BH field. The numerical code,
described in \cite{chen2013}, provides a high resolution shock
capturing calculation of the star as it approaches periastron and
follows the debris several dynamical timescales after the tidal
accelerations diminish.  Quantities in the BH frame are then obtained
from the local frame by series transformations given below.  In our
study, we use the sign conventions and notation of \cite{misn1973},
geometrical units in which $c = G = 1$, and scale all dimensional
quantities in the simulations relative to the BH mass $M$.  We use
Greek indices to label the four-dimensional coordinates and components
of tensors of the BH spacetime, latin indices for the FNC frame, and
reserve $i, j, k, l$ for FNC spatial coordinates.

\subsection{Frame trajectory and tidal potential}
For Newtonian encounters, the frame is evolved along a parametrized
parabolic orbit given in \cite{land1969}.  We use a tidal potential
derived from the multipole expansion of the Newtonian gravitational
potential $\Phi(r)$ for the BH.  The local frame is located
at a distance $\vec{R}_0$ from the BH.  The expansion of $\Phi$ at a
distance $\vec{x}$ from the origin of the local frame is
\ba\label{eqn:expansion_of_Newtonian_gravitational_potential} \Phi \nn
\nn & = & -\frac{GM}{R_0} + \frac{GM}{R_0^2}\vec{x}\cdot\vec{n} +
\tfrac{1}{2} C_{ij} x^i x^j + \tfrac{1}{6} C_{ijk} x^i x^j x^k \\
& & \qquad \qquad \qquad \qquad \ + \tfrac{1}{24} C_{ijkl} x^i x^j x^k x^l + \cdots, \ea
where $\vec{n}=\vec{x}/R_0$, with tidal tensor definitions $C_{ij}
\equiv \left ( \partial_i \partial_j \Phi \right )_0$ (quadrupole),
$C_{ijk} \equiv \left ( \partial_i \partial_j \partial_k \Phi \right
)_0$ (octupole), $C_{ijkl} \equiv \left ( \partial_i \partial_j
\partial_k \partial_l \Phi \right )_0$ (hexadecapole), etc., which
depend on the evolution of the local frame center along the
trajectory.  The gravitational acceleration due to the BH in this local
frame is $\partial_i\Phi$, where the first term is the frame
acceleration and subsequent terms are the tidal accelerations.  The
tidal potential is then defined as
\be\label{eqn:tidal_potential}
\Phi_{\rm tidal} = \tfrac{1}{2} C_{ij}
x^i x^j + \tfrac{1}{6} C_{ijk} x^i x^j x^k + \tfrac{1}{24} C_{ijkl} x^i x^j
x^k x^l + \cdots.
\ee 
We apply only the tidal accelerations $\partial_i\Phi_{\rm tidal}$ to
the self-gravitating star in the local frame.  For relativistic
encounters, the frame is evolved along a trajectory in the
Schwarzschild spacetime with parametrization given in \cite{cutl1994}.
We use a tidal potential as in Eq.~\eqref{eqn:tidal_potential},
derived from the formalism of \cite{ishi2005,chen2013}, with tidal
tensors $C_{ij} \equiv R_{0i0j}$, $C_{ijk} \equiv R_{0(i|0|j;k)}$, and
$C_{ijkl} \equiv R_{0(i|0|j;kl)}$ defined by the FNC Riemann tensor
$R_{abcd}$ and its derivatives with respect to FNC spatial
coordinates.  

\subsection{Validity of the FNC approximation}
\label{subsec:FNC_approx}


In the following, we show the validity of the FNC approximation
$g_{ab}$ of the BH spacetime $g_{\mu\nu}$ for all encounters in this
paper (listed below in Table \ref{tab:encounters_parameters})
\cite{chen2013}.  In order for this approximation to be valid
the size of the computational domain must be smaller
than the characteristic length scale of the tidal field, which is
smallest at periastron $R_p$.  In this study, the extent of the
computational domain is $\mathcal{L} = 16R_*$ and we have
$\mathcal{L}/R_p \ll 1$ for all encounters.  The correction due to the
motion of the BH (relative error of $\sim\mu$, where the largest mass
ratio is $\mu\sim 10^{-5}$) is neglected.  We account for the error in
neglecting terms higher than the hexadecapole term in the metric
expansion as well as the gravitomagnetic term.  We consider stars that
have stellar post-Newtonian (PN) velocity scale $\varepsilon^2 =
GM_*/(c^2R_*) \ll 1$.
%
%
Comparing the leading neglected term higher than the hexadecapole with
the Newtonian quadrupole, which is the dominant force term,
%
we obtain a fractional error in neglecting higher order quartic tidal
terms at a stellar radius $R_*$ from the origin of the FNC frame,
\be\label{eqn:fractional_error_quartic}
\mathcal{E}_{\rm quartic} \simeq \frac{M}{R_p^3} R_*^2 = \varepsilon^2 \eta^{-2}.
\ee
%
%
We consider the error in neglecting the gravitomagnetic potential by
comparing the leading term in $g_{0m}$ multiplied by the maximum
break-up velocity for the star $\varepsilon$ with the size of the
Newtonian quadrupole and obtain
%
%
%
%
\be\label{eqn:fractional_error_gravito-magnetic term}
\mathcal{E}_{\rm GM,max} \simeq \varepsilon^2 \mu^{-1/3} \eta^{-1/3}.
\ee
We find that the fractional error in neglecting terms above the
hexadecapole term is at most $\mathcal{E}_{\rm quartic}\sim 10^{-5}$
for main sequence stars and white dwarfs.  The fractional error in
neglecting the gravitomagnetic term is at most $\mathcal{E}_{\rm
  GM,max} \sim 10^{-4}$ for the main sequence star models and $\sim
10^{-3}$ for the white dwarf model.  Given the low level of error associated with 
the higher order terms in the FNC expansion, we assume that they can be safely neglected for all parameter choices in
this study.


\begin{table*}
\caption{\label{tab:encounters_parameters}
Parameters of initial polytropic models ($n=1.5$, $\gamma=5/3$)
for a low mass main sequence star (MS1, MS2, MS3: $M_* =
M_\odot$, $R_* = R_\odot$) and a white dwarf (WD: $M_* = 0.6 M_\odot$,
$R_* = 1.51\times 10^{-2} R_\odot$).  Note $R_* = \mu^{1/3} \eta^{-2/3} R_p$.
}
\begin{ruledtabular}
\begin{tabular}{cccccccccccc}
Label  & $\eta$  & $\beta$ & $M[M_\odot]$ & $R_p[M]$ & $R_p$[cm] & $\ell_0[M]$ & $\ell_0 [$cm$^2 $s$^{-1}]$ & $t_p[M]$ & $t_p$[s] & $P_m[M]$ & $P_m$[s]  \\
\hline
\hline
MS1 & 1.0   & 1.0 & 1.0e+05  & 2.2e+02 & 3.2e+12 & 21.0 (20.9) & 9.3e+21 & 3.2e+03 & 1.6e+03 & 2.3e+06 & 1.1e+06\\
MS2 & 1.0   & 1.0 & 1.0e+06 & 4.7e+01 & 6.7e+12 & 9.9 (9.71) & 4.4e+22 & 3.2e+02 & 1.6e+03 & 7.2e+05 & 3.5e+06\\
MS3 & 1.0   & 1.0   & 1.0e+07 &  1.0e+01 & 1.5e+13 & 5.0 (4.51) & 2.2e+23& 3.2e+01 & 1.6e+03 & 2.3e+05 & 1.1e+07\\
WD5 & 1.44 & 0.784 & 1.0e+05 & 5.0  & 7.4e+10 & 4.1 (3.16) & 1.8e+21 & 1.1e+01 & 5.5e+00 & 1.5e+04 & 7.2e+03\\
WD6 & 1.89 & 0.654 & 1.0e+05   & 6.0  & 8.9e+10 & 4.2 (3.46) & 1.9e+21 & 1.5e+01 & 7.2e+00 & 2.5e+04 & 1.2e+04\\
\end{tabular}
\end{ruledtabular}
\end{table*}

\begin{figure*}
{ \begin{center} 
\includegraphics[scale=1.1]{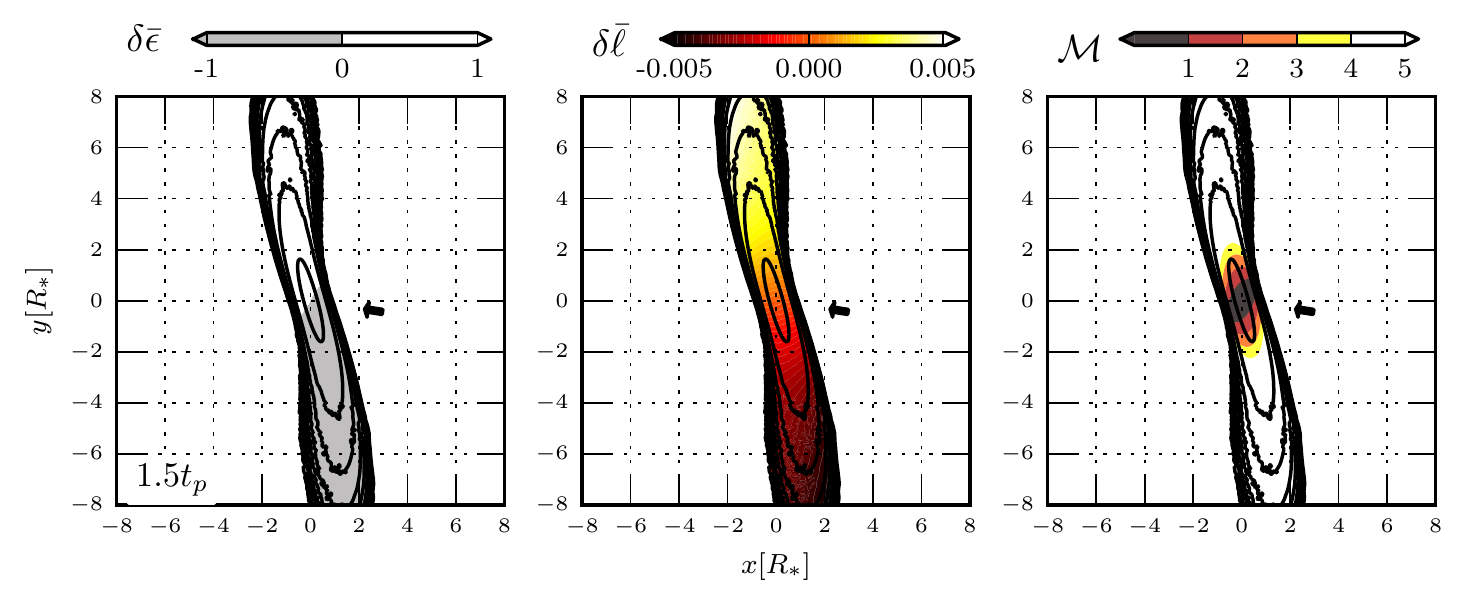} 
\end{center} 
\caption{\label{figure:contours}
Normalized distribution of the specific orbital energy $\epsilon$ (left), specific
orbital angular momentum $\ell$ (middle), and the Mach number $\mathcal{M}$ (right panel) of the debris at 1.5$t_p$ after the periastron passage for relativistic model MS3.  The density contours (black lines) are of $\log_{10}
(\rho/\rho_c )$, ranging from -5 to 0 in steps of 0.5,  where $\rho_c$ is the initial central density of the
star.  Arrow indicates the direction from the BH to the star.
%
}  }
\end{figure*}

\subsection{Transformation from local to BH frame}
\label{subsec:transformation}
We transform quantities calculated in the local frame to the black
hole frame in order to obtain the orbital parameters of the debris.
For the Newtonian case, this is a straightforward Galilean
transformation.  For the relativistic case, quantities corresponding
to an event $x^a = (\tau,x^i)$ in the FNC frame are transformed to the
BH frame at coordinate location $X^{\mu}$ by series expansion about
the origin of the local moving frame, $X^{\mu}_{(0)}$.  This is
possible because the positions and velocities of the fluid elements in
the local frame are smaller than the position and motion of the frame
with respect to the BH.  For the fluid velocities considered in this
paper, we include more terms in the expansion than \cite{chen2013}.
For the coordinate transformation, we form a Taylor expansion in the
spatial coordinates $x^i$,
\ba\label{eqn:FNC_to_BH_coordinate_transformation}
X^{\mu}(\tau,x^i)
\nn & = & X^{\mu}_{(0)}(\tau) + \lambda_i^{\ \mu}x^i + \tfrac{1}{2} \sigma_{ij}^{\ \ \mu}x^i x^j \\
& & \qquad \quad \ + \tfrac{1}{6} \kappa_{ijk} x^i x^j x^k 
+ \cdots,
\ea
where the coefficients $\lambda^{\ \mu}_{i}$, $\sigma^{\ \ \mu}_{ij}$,
and $\kappa^{\ \ \ \mu}_{ijk}$
 are evaluated at $X_{(0)}^{\mu}$.  For the velocity transformation,
 we expand about the frame center in both spatial coordinates $x^i$
 and velocities $v^i$.  We write the four-velocity in the BH frame as
 $U^{\mu} = dX^\mu/d\tau = (\partial X^{\mu}/\partial \tau) \, u^0 +
 (\partial X^{\mu}/\partial x^i) \, u^i$, where in the FNC frame the
 four-velocity reduces to $u^0 \simeq 1 + \mathcal{O}(\varepsilon^2)$
 and $u^i = v^i + \mathcal{O}(\varepsilon^3)$ in neglecting
 corrections at or below the size of the stellar PN order
 $\varepsilon^2$.  Expanding $X^\mu$, we have
\ba\label{eqn:FNC_to_BH_velocity_transformation}
U^{\mu}(\tau,x^i,v^i) 
\nn & = &
U^{\mu}_{(0)} + 
(\tfrac{d}{d\tau} \lambda_i^{\ \mu}) x^i
+ \tfrac{1}{2}(\tfrac{d}{d\tau} \sigma_{ij}^{\ \ \mu}) x^i x^j\\
\nn & + & 
\tfrac{1}{6}(\tfrac{d}{d\tau} \kappa_{ijk}^{\ \ \ \mu}) x^i x^j x^k
\lambda_i^{\ \mu} v^i 
+ \sigma_{ij}^{\ \ \mu} v^i x^j \\
 & + &
\tfrac{1}{2}\kappa_{ijk}^{\ \ \ \mu} v^i x^j x^k
+ \cdots,
\ea
with coefficients evaluated at $X^\mu_{(0)}$.  We have that
$\lambda_{i}^{\ \mu}$ is the FNC tetrad \cite{chen2013} and obtain
$\sigma_{ij}^{\ \ \mu}$ and $\kappa_{ijk}^{ \ \ \ \mu}$ in terms of
the connection coefficients $\Gamma^\mu_{\alpha\beta}$ of the BH
spacetime in Appendix \ref{sec:expansion}.  For the encounters in this
paper, higher order terms (than those mentioned above) in the
expansion are smaller than the overall relative error in the numerical
method, $\mathcal{E}\lesssim 10^{-3}$ (see Section
\ref{subsec:FNC_approx}), and are neglected.  Using the position
$X^\mu$ and velocity $U^\mu$ of a debris element in the BH frame, we
calculate the orbital parameters represented by specific orbital energy
$\epsilon$ and specific orbital angular momentum $\ell$ (or alternatively periastron
distance $R_p$ and eccentricity $e$).  We outline this calculation in
Appendix \ref{sec:orb_param}.


\begin{figure*}
{ \begin{center} 
\includegraphics[scale=1.1]{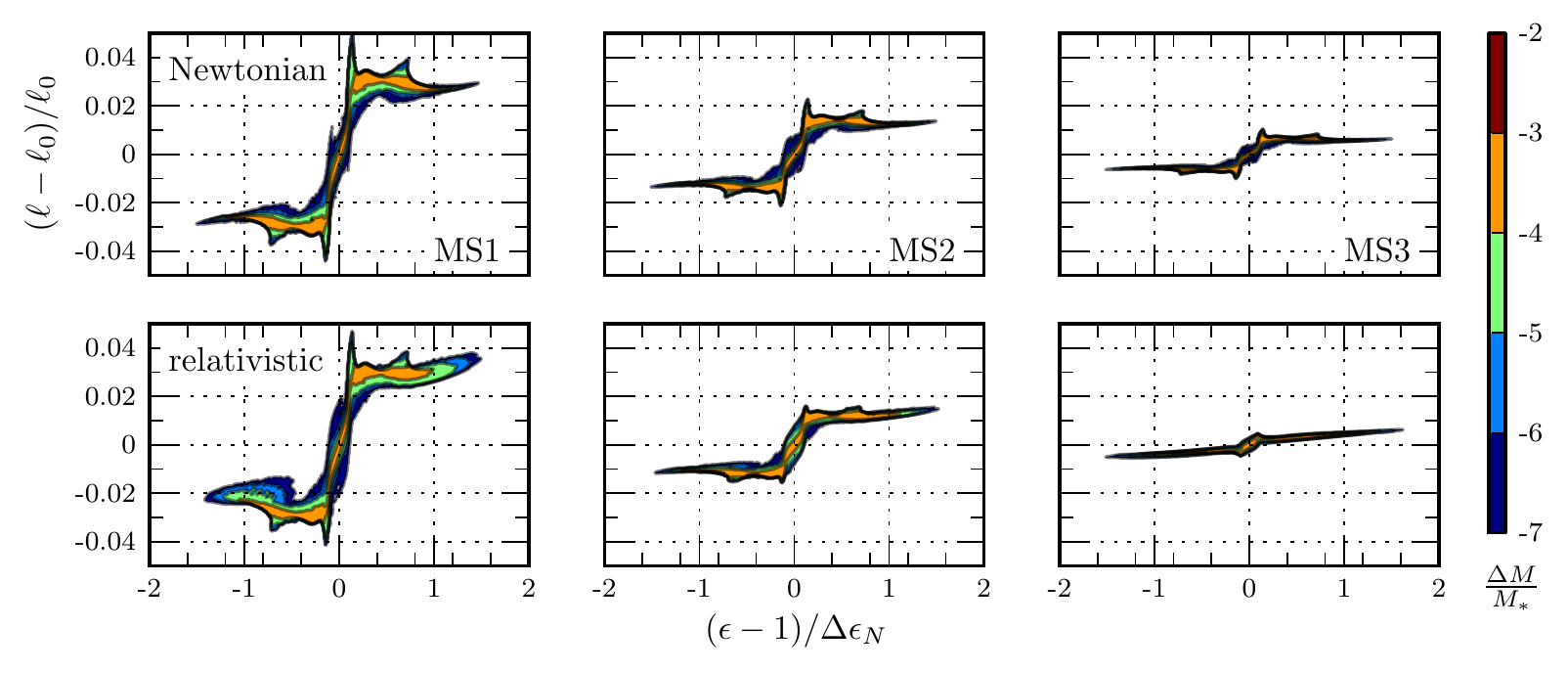} 
\end{center} 
\caption{\label{figure:mass_contours_l_v_e_MS} 
Distribution of specific
  orbital energy and angular momentum of the debris at the end of the
  encounter for Newtonian and relativistic models MS1, MS2, and MS3. Color indicates the stellar mass fraction enclosed in every contour level.
} }
\end{figure*}

\section{Results}
\label{sec:results}

\subsection{Initial setup}

In this study, we consider $n=3/2$ polytropes with $\gamma=5/3$
adiabatic index for modeling a low mass main sequence star (MS1, MS2,
MS3: $M_* = M_\odot$, $R_* = R_\odot$) and a white dwarf (WD5, WD6:
$M_* = 0.6 M_\odot$, $R_* = 1.51\times 10^{-2} R_\odot$).  We simulate
$\sim 20 t_p$ before and $\sim 40 t_p$ after periastron passage.  In
Table \ref{tab:encounters_parameters}, we give the parameters for the
initial models simulated in this paper in units of geometrized BH mass
$M$ and CGS units.  We give encounter strength $\eta$ in
Eq. \eqref{eqn:encounter_strength}, the corresponding penetration
factor $\beta$, mass of the BH $M$, periastron $R_p$, relativistic
specific angular momentum of the frame trajectory (Newtonian quantity
in parentheses) $\ell_0$, orbital timescale at periastron $t_p$, and
estimated Keplerian return period of the most bound debris $P_m$ in
Eq. \eqref{eqn:Newtonian_estimated_accretion_rate}.

The star is centered in a computational domain of length
$\mathcal{L}=16R_*$ in order to follow the evolution of the remnant
core and tidal debris streams as the stellar fluid expands in the
encounter.  Each simulation is modeled using a uniform grid with a
resolution of $32$ zones per stellar radius $R_*$.  The numerical
accuracy in computing the energy and angular momentum deposited onto
the star with this code at different resolutions is discussed in
\cite{chen2013}. In particular, convergence is shown in simulations of
tidally transferred energy (Fig. 6) with angular momentum conservation
(Fig. 15). Resolution higher than 32 zones per radius is necessary for
weak tidal encounters $\eta=4$. For encounters with greater energy and
angular momentum transfer $\eta\gtrsim 2$, convergence is achieved
with lower levels of resolution. In this paper, we model encounters of
strength $\eta \sim 1$, at the threshold of disruption.

\subsection{Orbital dynamics of the debris}

Orbital dynamics of the debris in the BH frame is characterized
by $\epsilon$ and $\ell$.  
The left and center panel of Fig.~\ref{figure:contours} show the
normalized change in specific orbital energy $\epsilon$ and angular
momentum $\ell$ of the debris, respectively, at $1.5t_p$ after
periastron passage.  We normalize the distribution in specific orbital
energy as $\delta\bar\epsilon = (\epsilon - \epsilon_0)/|\epsilon|$, where
$\epsilon_0=0$ (or $\epsilon_0=1$ for the relativistic case) is the
specific orbital energy of the trajectory of the frame. Similarly, the
normalized distribution in specific angular momentum is calculated as
$\delta\bar\ell = (\ell - \ell_0)/\ell_0$, where $\ell_0$ is the specific
orbital angular momentum of the initial trajectory of the frame. The
left and center panel of Fig.~\ref{figure:contours} show that the
tidal lobe bound to the BH has less angular momentum than the initial
trajectory as well as a gradient in the distribution of angular
momentum. This implies that the expanding debris will eventually
occupy a range of orbits with different semi-major axes and
eccentricities.
 
The right panel of Fig.~\ref{figure:contours} is a contour plot of the
Mach number $\mathcal{M} = |\vec{v}|/c_s$ for the locally defined
fluid velocity $\vec{v}$ and sound speed $c_s$.  The remnant core of
the star is subsonic $\mathcal{M} <1$ and the two tidal debris streams
are supersonic, in agreement with the findings of \cite{frol1994}. We
track the tidal lobes as they escape the domain and include the
supersonic material (which is no longer causally connected to the
remnant core hydrodynamically) in our estimate of the orbital
parameters of the debris.

\begin{figure*}
{ \begin{center} 
\begin{tabular}{cc} 
\includegraphics[scale=0.95]{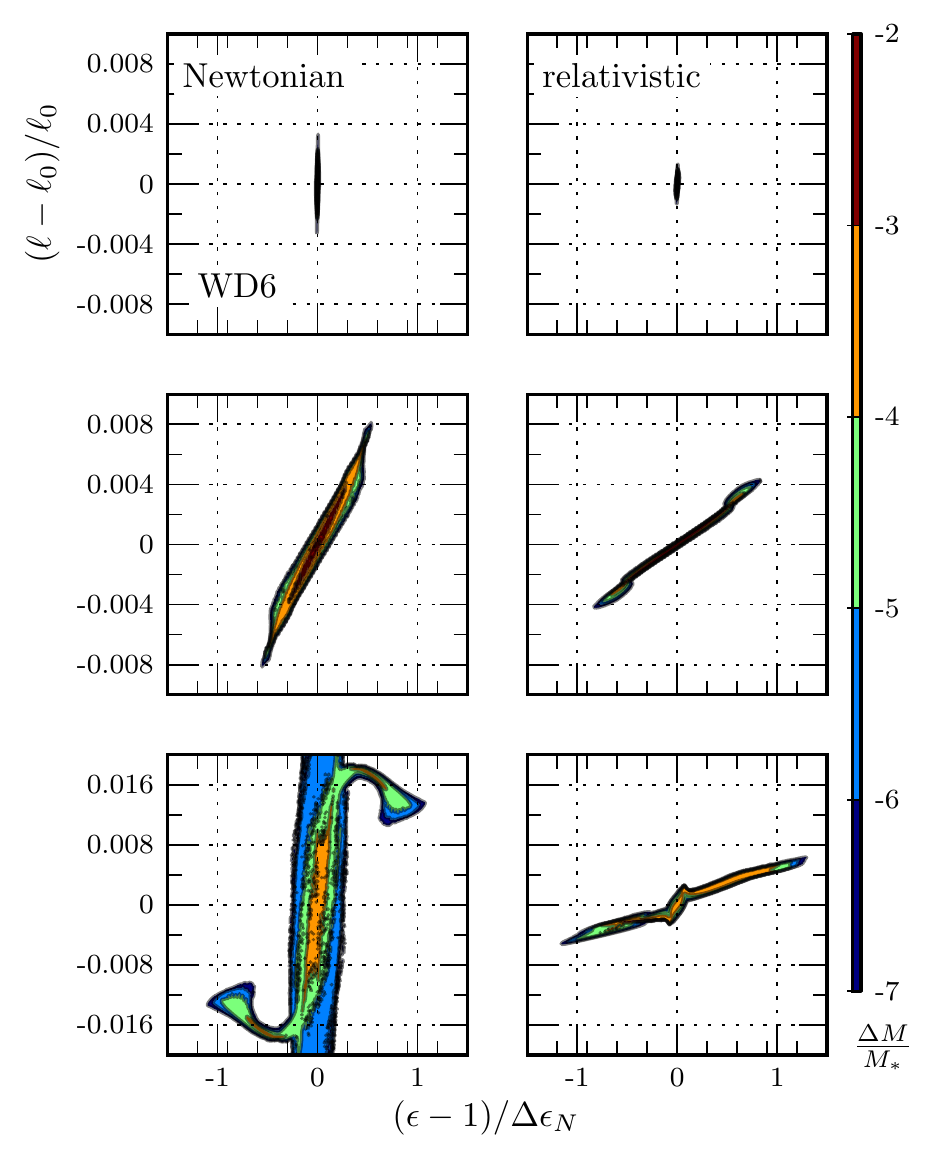} &
\includegraphics[scale=0.95]{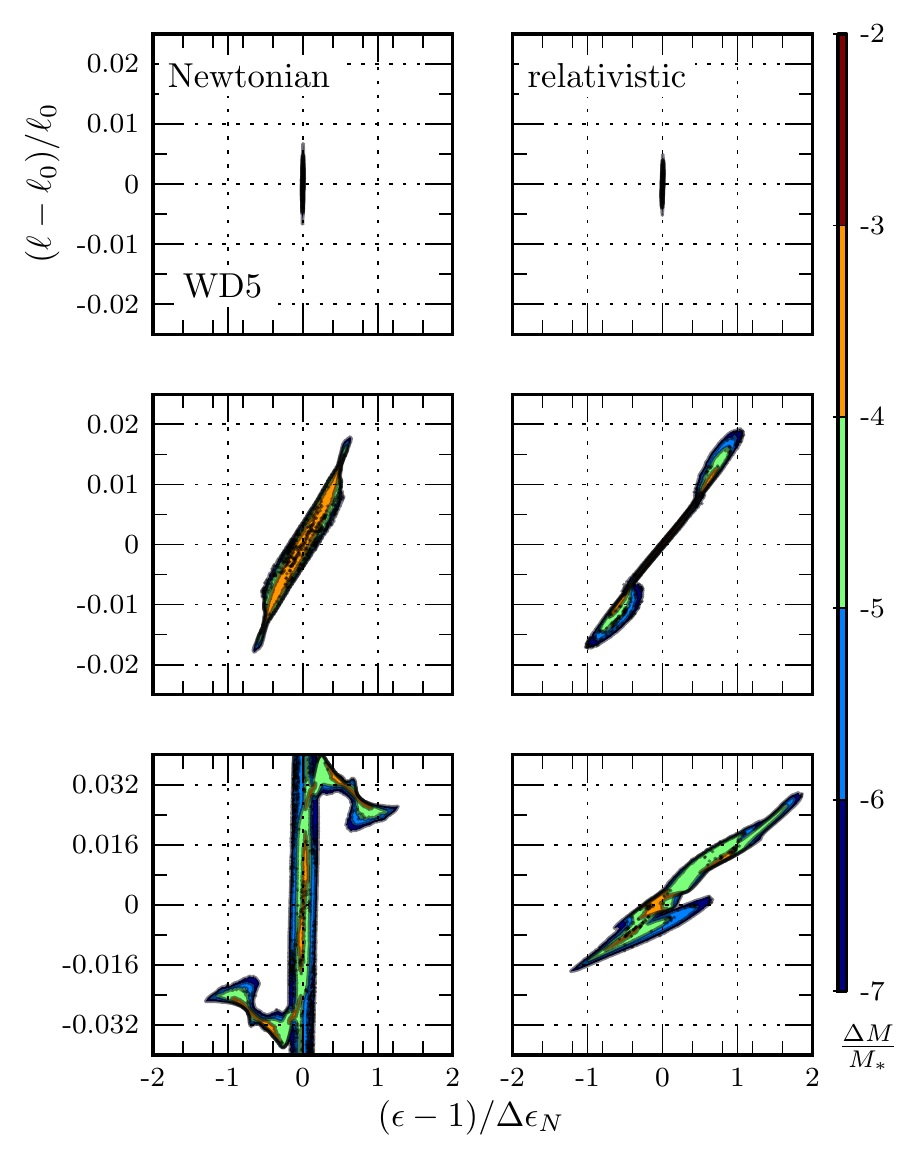} \\
\end{tabular}

\end{center} 
\caption{\label{figure:mass_contours_l_v_e_WD} Distribution of
  specific orbital energy and angular momentum of the debris for
  Newtonian and relativistic model WD6 (left) and model WD5 (right) at
  different points in the encounter: initially (top row), periastron
  (middle row), end of encounter (bottom row).  }}
\end{figure*}

\begin{figure*}
{ \begin{center} 
\includegraphics[scale=0.8]{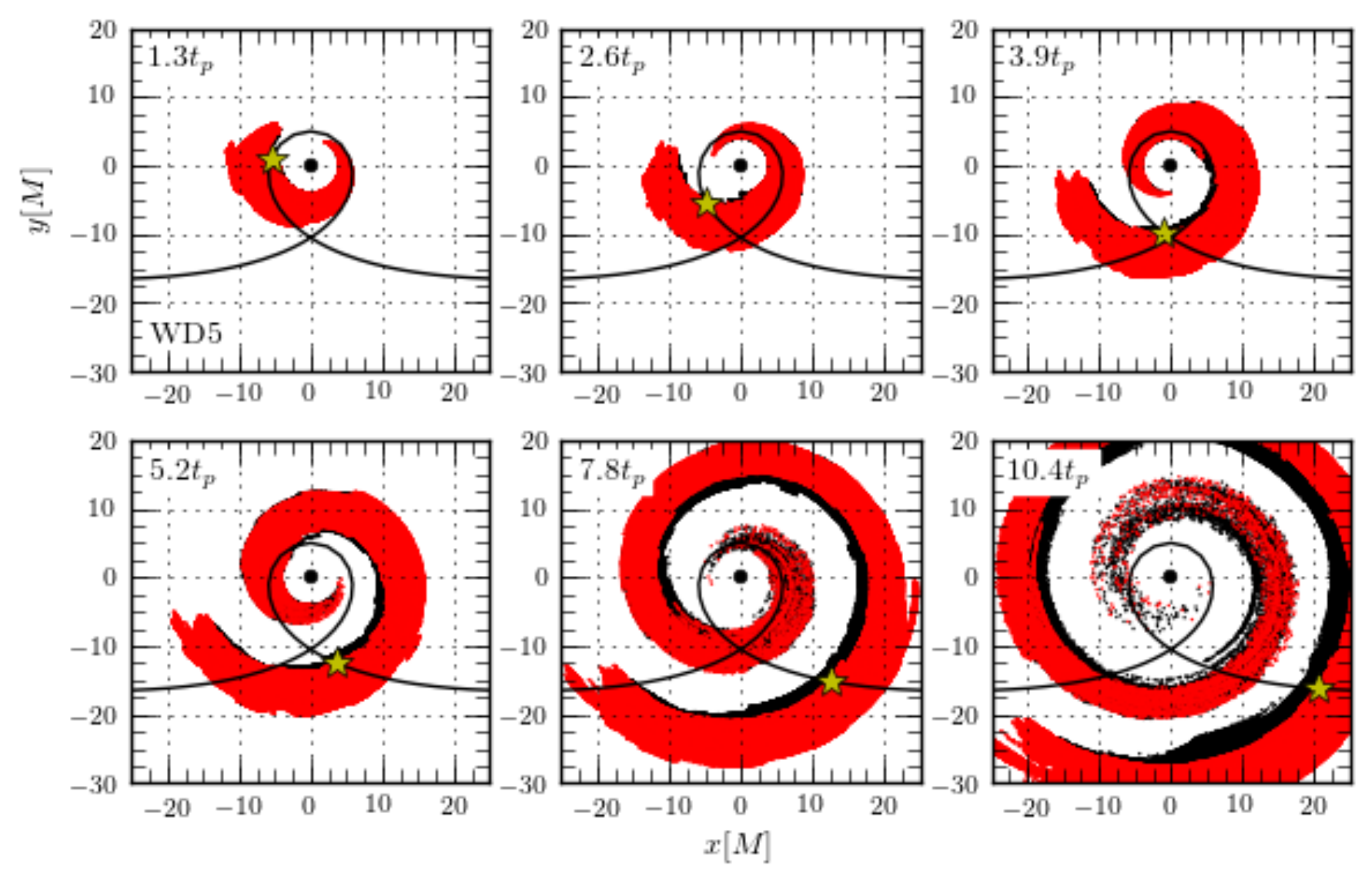}
\end{center} 
\caption{\label{figure:BH_frame_WD5} Debris streams in the BH frame
  for relativistic model WD5.  Black (red) color marks a portion of
  the debris stream bound (unbound) to the BH.  Note that the streams
  are overlapping with 45\% of the debris bound to the black hole.
  Star marker follows the initial trajectory of the star.  }}
\end{figure*}

\subsection{Distribution of specific orbital energy and angular momentum}
\label{sec:relativistic_effects_distribution}

We construct the maps of the distribution of $\epsilon$ and $\ell$ of
the debris for all tidal encounters studied in this paper and
investigate their evolution with BH mass and stellar type. In
Fig.~\ref{figure:mass_contours_l_v_e_MS}, $\epsilon$ and $\ell$ are
given at the end of Newtonian and relativistic simulations of models
MS1, MS2, and MS3. The contour levels indicate the enclosed fraction
of the debris mass, $\Delta M/M_*$. In this figure, the change in
specific orbital energy is normalized by the Newtonian spread in
energy $\Delta\epsilon_N$ in
Eq. \eqref{eqn:Newtonian_spread_of_energy} and the normalized change
in specific orbital angular momentum is the same as
Fig.~\ref{figure:contours}.  In
Fig.~\ref{figure:mass_contours_l_v_e_MS}, the spread in $\ell$
decreases with the BH mass for both the Newtonian and relativistic
simulations.  The spread in $\epsilon$ is similar for both and larger
than $2\Delta\epsilon_N$, indicating a full disruption of the star.

Comparing the Newtonian and relativistic simulations in
Fig.~\ref{figure:mass_contours_l_v_e_MS}, we find a further decrease
in the width of the $\ell$-distribution for relativistic cases
relative to their Newtonian counterparts. This effect is even more
pronounced in models WD6 and WD5, shown in the bottom rows of
Fig.~\ref{figure:mass_contours_l_v_e_WD}.  The narrowing of the
angular momentum distribution in relativistic simulations for models
MS3, WD5 and WD6 indicates that in these simulations debris is
confined to a narrower range of eccentric orbits. Since only the
$\epsilon < 0$ branch remains bound to the BH, for which
$\delta\bar\ell\lesssim0$, it follows that all bound orbits will be
highly eccentric relative to the initial orbit of the incoming star
and that the Newtonian simulations give rise to more eccentric debris
orbits than relativistic ones. As relativistic simulations are similar
to their Newtonian equivalents in every aspect except in the
relativistic treatment of the tidal interaction with the BH, it
follows that this effect is purely relativistic.

In Fig.~\ref{figure:mass_contours_l_v_e_WD}, we show the evolution of
the $\epsilon-\ell$ distribution with time for WD6 and WD5 (nominally
$\eta= 1.89$ and $1.44$ encounters, respectively) at the beginning of
the encounter (top row), periastron (middle row), and the end of the
encounter (bottom row).  For both the Newtonian and relativistic
simulations, the distribution in $\epsilon$ and $\ell$ initially
broadens as the star passes through periastron.  

In Newtonian simulations, the distribution in energy after periastron
passage evolves to become more ``vertical". This is an indication of the
diminishing tidal accelerations and increasing significance of the
self-gravity of the stellar remnant, which in this case was not fully
disrupted as it continues to recede from the BH. This effect is also
seen in simulations of disruptions by
\cite{loda2009,haya2013,guil2013} and is manifested as a central spike
in their one-dimensional plots of the distribution of mass and energy
$dM/d\epsilon$.  Note that similar 1D distributions can be obtained
from our $\epsilon-\ell$ maps by summing along the angular momentum
(vertical) axis.

After periastron passage, the distribution in orbital energy of the
debris is wider in relativistic simulations. Unlike in the Newtonian
simulations, the spread in energy at periastron in relativistic
simulations indicates full disruption. This is consistent with
findings of \cite{frol1994,kesd2012}, who note that differences
between the relativistic and Newtonian tidal tensor imply that the
relativistic tidal interactions are more disruptive.


An additional feature of the $\epsilon-\ell$ maps, noticeable in the
most relativistic encounter WD5
(Fig.~\ref{figure:mass_contours_l_v_e_WD}), is that the symmetry in
the distribution of the debris breaks down after periastron passage
(bottom right panel). Specifically, the tidal lobe bound to the BH
evolves to be less strongly bound than it would otherwise be if the
symmetry in the distribution was preserved as in the WD6 model. We show
this to be a consequence of apsidal precession in the next section,
where we describe the dynamics of the debris streams in the black hole
frame.

Using the orbital parameters of the debris
calculated from the maps of $\epsilon$ and $\ell$ (shown in
Figs.~\ref{figure:mass_contours_l_v_e_MS} and
\ref{figure:mass_contours_l_v_e_WD}), we construct {\it ballistic
  debris elements} from a collection of fluid elements that share
similar values in $\epsilon$ and $\ell$.  We then propagate them
ballistically in time according to the procedures for the Newtonian
and Schwarzschild spacetimes described in Appendix
\ref{sec:orb_param}.

\begin{figure*}
{ \begin{center} 
\includegraphics[scale=0.925]{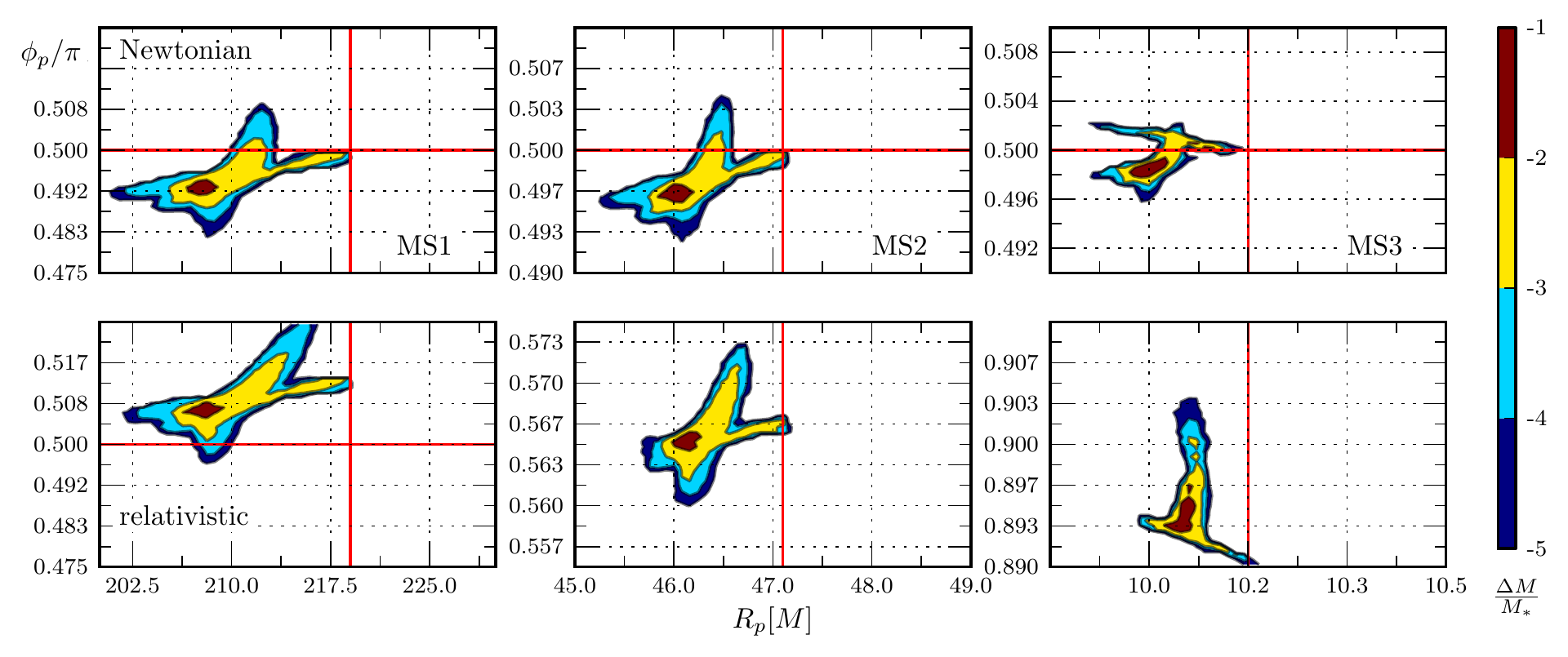} 
\end{center} 
\caption{\label{figure:periastron_advance_MS} Periastron location of the debris for Newtonian and relativistic
  simulations MS1, MS2, and MS3.  Periastron location of the initial orbit is indicated by the cross hairs.} }
\end{figure*}

\begin{figure}
{ \begin{center} 
\includegraphics[scale=.75]{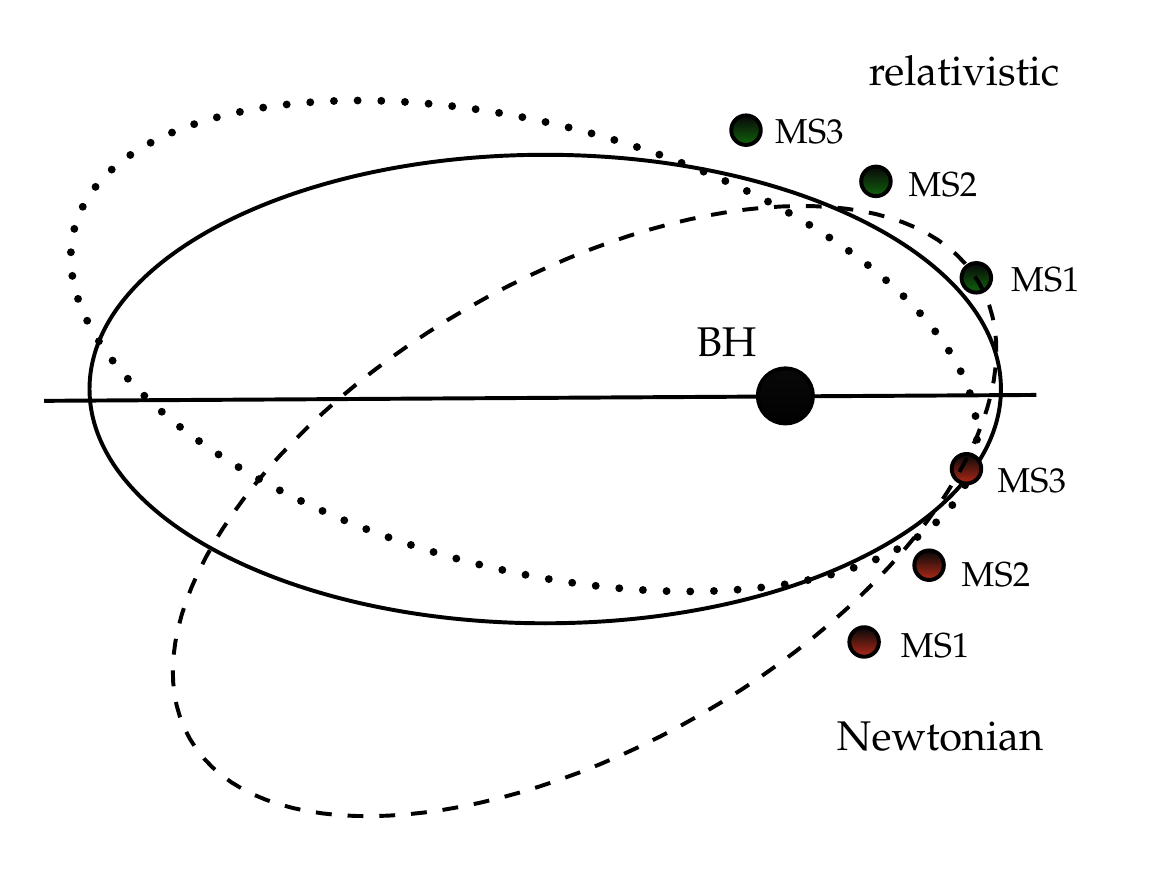} 
\end{center} 
\caption{\label{figure:periastron_advance} Illustration of azimuthal locations of periastrons for orbits of bound debris elements after disruption (not to scale).  In Newtonian simulations debris arrives to periastrons with the offset of $\lesssim 2\pi$ in azimuth (dotted line) relative to the periastron of the initial orbit (solid). In relativistic simulations debris arrives to periastrons with the offset of $> 2\pi$ in azimuth due to the apsidal advance of the orbit (dashed). The initial orbit is counter-clockwise.}}
\end{figure}
\subsection{Whirling debris streams in the black hole frame}



In this section, we report a mixing of bound/unbound streams for the
most relativistic encounter WD5.  In a Newtonian simulation, the
debris streams should be distinct (opposite sides of the remnant core)
with roughly 50\% of the debris bound to the black hole.  For the
relativistic WD5 encounter, we find that 45\% of the debris is bound
to the black hole.  In Fig.~\ref{figure:BH_frame_WD5}, we show the
evolution of the ballistic debris elements in the relativistic run
WD5. We also denote the location of a test particle that follows the
initial orbit of the star. In this plot, the black debris stream
represents material bound to the BH while the red debris stream is
unbound.  In this encounter the apsidal precession shapes the
distribution of the debris and the result is a whirling, crescent
shaped tidal tail where bound and unbound material overlap and
mix. The same effects are present in the run WD6 to a lesser
extent. This distribution and mixing of the bound and unbound debris
share resemblance with the relativistic MS-BH disruptions by
\cite{koba2004} as well as neutron star - BH disruptions by
\cite{etie2008,robe2011,fouc2012,east2012,kyut2013}. They indicate
that the orbital precession is likely to lead to orbit intersection
and collisions of fluid elements as they pass through the
periastron. This differs from the Newtonian encounters, where the
initial orbital trajectory clearly delineates the bound and unbound
debris streams.  We caution that we evolve the debris ballistically
and therefore do not account for hydrodynamic interactions and
self-gravity. This approximation ceases to be valid once the orbits of
the fluid elements intersect.  Therefore, a full hydrodynamic
treatment is necessary in order to follow the evolution of the debris
accretion disk.


\begin{figure*}
{ \begin{center}
\begin{tabular}{cc} 
\includegraphics[scale=1.0]{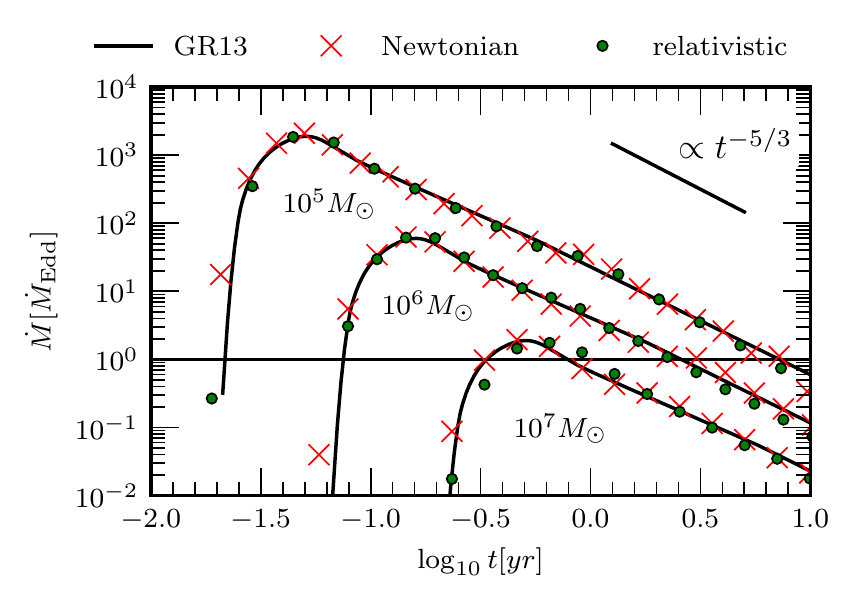} &
\includegraphics[scale=1.0]{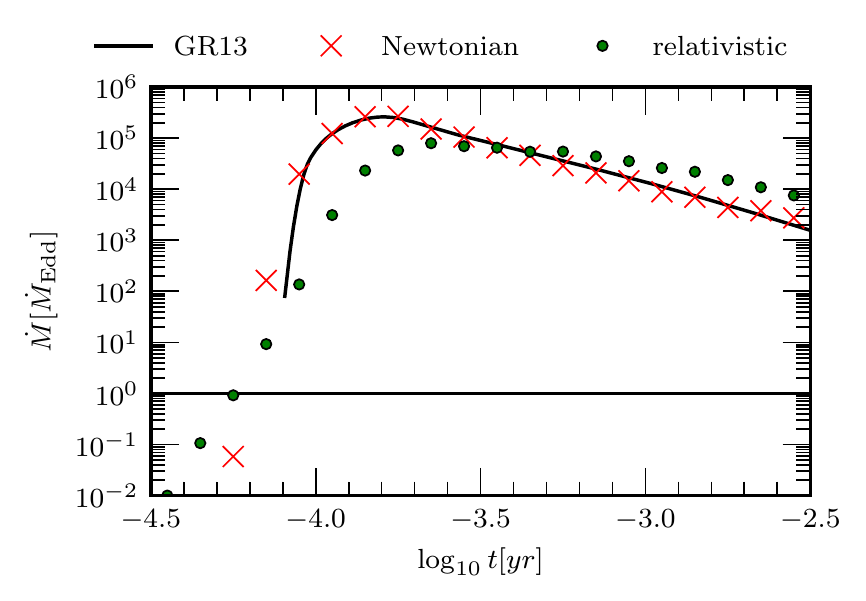} \\
\end{tabular}
\end{center} 
\caption{\label{figure:return_rates} Rates of return of the debris to periastrons as a function of time for MS1, MS2, MS3 ({\it left})
and WD models ({\it right}). Rates from Newtonian (red $\times$) and relativistic simulations (green circles) are compared with parametric fits from \cite{guil2013} (solid black line).}}
\end{figure*}

\subsection{Apsidal advance in the debris orbits}
\label{subsec:apsidal}
In order to investigate the relativistic effects which result in an
apsidal advance of the debris orbits, we map the periastron locations
of the debris in azimuthal angle ($\phi_p$) and radius ($R_p$) for
both Newtonian and relativistic simulations
(Fig.~\ref{figure:periastron_advance_MS}).  Both sets of simulations
show that the debris returns to the periastron radii that are slightly
smaller than that of the original orbit (indicated by the vertical red
line). They however exhibit different behaviors when it comes to the
distribution of the debris in the azimuthal angle.

In Newtonian simulations, following the disruption, the bulk of the
debris returns to the periastron in $\lesssim2\pi$ azimuthal rotation
relative to the initial periastron located at $\pi/2$. This can be
understood in terms of the tidal interaction between a star and a BH.
If the disruption of a star occurred instantaneously at periastron,
the debris would return to this point on the orbit in a full $2\pi$
azimuthal rotation.  In reality however, the star is starting to be
tidally deformed and spun up before it reaches periastron and
consequently, its fluid elements start acquiring a range of values in
angular momentum. The effect of azimuthal offset in Newtonian
simulations decreases with the BH mass; it is strongest in the MS1 and
weakest in the MS3 run (see Fig.~\ref{figure:periastron_advance} for
illustration of the periastron locations for the bulk of the stellar
debris in different simulations). Note that this is consistent with
the narrowing in the angular momentum distribution of the debris with
the BH mass shown in Fig.~\ref{figure:mass_contours_l_v_e_MS}.

\begin{table*}
\caption{\label{tab:return_rates}
Comparison of Newtonian and relativistic simulations.  Peak return
rate ($\dot{M}_{\rm peak}$), time when $\dot{M}=\dot{M}_{\rm Edd}$ ($t_{\rm Edd}$),
time when $\dot{M}=\dot{M}_{\rm peak}$ ($t_{\rm peak}$), delay in peak time ($\Delta t_{\rm
  delay} = t^{R}_{\rm peak} - t_{\rm peak}^N$), power law index for the return rate from
relativistic simulations ($n_{\infty}^R$), and estimated BH mass ($M_{\rm est}$).
}
\begin{ruledtabular}
\begin{tabular}{cccccccccc}
Label  &$\dot{M}^N_{\rm peak}[\dot{M}_{\rm Edd}]$ & $\dot{M}^R_{\rm peak}[\dot{M}_{\rm Edd}]$ & $t^N_{\rm Edd}$[days] & $t^R_{\rm Edd}$[days] & $t^N_{\rm peak}$[days] & $t^R_{\rm peak}$[days] & $\Delta t_{\rm delay}$[days] & $n_{\infty}^R$  & $M_{\rm est}[M_\odot]$ \\
\hline
\hline
MS1    & 2.1e+03 & 2.2e+03 & 7.1 & 7.6 & 1.9e+01 & 1.9e+01 & 0.0  & -1.64  & 1.00e+05  \\
MS2    & 6.9e+01  & 7.0e+01 & 2.6e+01  & 2.8e+01 & 5.9e+01 & 6.5e+01 & 5.7  & -1.66  & 1.17e+06  \\
MS3   & 2.2  & 1.8 & 1.2e+02 & 1.4e+02  & 1.9e+02 & 2.1e+02 & 2.3e+01 & -1.66 & 1.26e+07   \\
WD5   & 2.9e+05 & 8.9e+04 & 2.3e-02 & 1.8e-02 & 5.9e-02 & 8.8e-02 & 2.8e-02  & -1.61 & 3.10e+05  \\
WD6    & 1.1e+05 & 2.9e+05 & 2.9e-02 & 2.9e-02 & 6.3e-02 & 1.0e-01 & 3.7e-02 & -1.73 & 3.06e+05  \\
\end{tabular}
\end{ruledtabular}
\end{table*}

In relativistic simulations the debris exhibits a positive shift in azimuthal angle relative to the Newtonian analogs which arises after $>2\pi$ azimuthal rotation due to the apsidal advance of the orbit (Fig.~\ref{figure:periastron_advance}). We compare the shifts measured from the MS1, MS2, and MS3 models with the orbital (1PN) post-Newtonian correction for apsidal advance,
\be\label{eqn:PN_apsidal_advance}
\Delta\phi_p^{\rm PN} = 6\pi \frac{GM}{c^2p},
\ee
\cite{wein1972}
where $p=R_p(1+e) \sim 2 R_p$.  From
Fig.~\ref{figure:periastron_advance_MS} we inspect the approximate
periastron radii occupied by the bulk of the stellar debris (contours
at the level $\Delta M/M_*\sim 10^{-1}$) in Newtonian simulations MS1,
MS2, and MS3 as $R_p\sim 208M, 46M, 10M$, respectively.  For these
values of periastron radii the 1PN correction estimates apsidal
advance of $\Delta\phi_p^{\rm PN} \sim 0.014\pi, 0.065\pi,
0.300\pi$. This is consistent with the shift measured between the
debris in the Newtonian and relativistic simulations of $0.014\pi$ for
MS1 (from $0.492\pi$ to $0.506\pi$), $0.068\pi$ for MS2 (from
$0.497\pi$ to $0.565\pi$) and $0.396\pi$ for MS3 (from $0.498\pi$ to
$0.894\pi$).  In comparing the Newtonian and relativistic result, we
note a departure from the orbital 1PN correction for the apsidal
advance of the bulk of debris as well as differences in the shape of
the distribution for increasingly relativistic encounters.  This
indicates that the combined effects of the tide and star's trajectory
contribute significantly to the apsidal advance in the relativistic
encounters as well.

\subsection{Relativistic effects in the return rate of debris}

To calculate the rate of return of debris to the new periastrons, we
construct a histogram in arrival time.  In
Fig.~\ref{figure:return_rates}, we show the return rates for models
MS1, MS2, MS3, and WD5 in units of $\dot{M}_{\rm Edd}$, where the
Eddington luminosity is $L_{\rm Edd} = 1.3 \times 10^{38}\;{\rm
  erg\,s^{-1}} (M/M_{\odot})$ and $\dot{M}_{\rm Edd} = L_{\rm
  Edd}/(0.1c^2)$ \cite{shap1983}.  We compare results of Newtonian and
relativistic simulations with fits based on Newtonian simulations by
\cite{guil2013} (solid black line).  In the left panel, we consider
MS1, MS2, and MS3 encounters at the threshold of disruption ($\eta=1$)
and compare with $\beta=1$ fits for disruption by $10^5, 10^6, 10^7
M_\odot$ BH.  In the right panel, we consider a partial disruption of
a white dwarf at $\eta=1.44, \beta=0.784$ and compare it with a
$\beta=0.8$ fit.

The return rate curves in all cases exhibit characteristic features
similar to previous works that include a rise to the peak followed by
a power law fall-off. Closer inspection and comparison with the fits
of \cite{guil2013} also reveal differences, which are somewhat
visually attenuated due to the nature of the log-log plot.  Newtonian
rates exhibit a slight earlier rise than the fits (GR13) while
relativistic rates show a more gradual rise to the peak.  We interpret
the differences between the Newtonian and relativistic rates as the
result of the apsidal advance of the debris orbits from a combination
of effects discussed in Sec.~\ref{subsec:apsidal}.

We note that once debris leaves the computational domain, we are
unable to account for the combined gravitational and hydrodynamical
interaction with what remains in the computational domain.  While this
is the case, we note that this interaction between the streams and the
core affects the least bound debris (which show up at later times in
the return rate) more than the debris that is initially and
supersonically stripped from the star close to periastron (which show
up at early times in the return rate).  We are also unable to account
for the self-gravity of the tidal streams themselves.  Nevertheless,
we compare our estimate of the return rate with \cite{guil2013}, who
simulate the interaction between the core and the streams until the
energy distribution reaches fixed values.  The differences in
Fig.~\ref{figure:return_rates} between our Newtonian simulations and
those by \cite{guil2013} reflect the differences in how the orbital
parameters of the debris are obtained from simulation as well as the
evolution of the ballistic debris orbits.  We note that these are
small in comparison to the differences between the relativistic and
Newtonian simulations.

For all simulations we measure the peak accretion rate ($\dot{M}_{\rm
  peak}$), the time when it occurs ($t_{\rm peak}$), as well as the
power-law fall-off ($n^N_{\infty}$) and find that the values from our
Newtonian simulations match well with the predictions of
\cite{guil2013}. With the exception of the MS1 run, where relativistic
effects are negligible, all other relativistic simulations exhibit
noticeable differences with respect to their Newtonian
counterparts. These are listed in Table~\ref{tab:return_rates} which
in addition to already mentioned parameters shows the time for the
return rate to reach the Eddington limit $t_{\rm Edd}$ and the shift
in peak time $\Delta t_{\rm delay}$ between the Newtonian and
relativistic simulations in days.

Table~\ref{tab:return_rates} and the right panel of
Fig.~\ref{figure:return_rates} show that the return rate in
relativistic simulation WD5, where apsidal advance is more severe,
is suppressed by a factor of $\sim 3$ relative to the Newtonian rate.
Along similar lines, the return rate curve in relativistic simulations
is broadened and the occurrence of the peak in the return rate
delayed. We attribute this delay to the apsidal advance of the
orbits and the property of relativistic encounters to confine the
angular momentum distribution of the debris to a relatively narrow
range of values. For example, we measure $\Delta t_{\rm delay}$ for
the MS2 and MS3 models of about 6 and 23 days, respectively. For the
WD5 and WD6 models, which play out on much shorter timescales, the
delays are only about $\sim 0.03-0.04$ days. In the next section we
discuss implications of the relativistic effects in the return rate
curves in the context of observations and estimates of the BH mass.

\section{Discussion}
\label{sec:discussion}
Our simulations show that relativistic effects manifest in the return
rate of the debris in two ways: they are noticeable in a more gradual
rise of the return rate curve to the peak as well as in the delay of
the peak rate relative to the Newtonian simulations.  As expected, the
magnitude of these effects increases for encounters that occur closer
to the BH and while they are negligible in the MS1 model, they are
present in MS2 and MS3 and fairly pronounced in the WD encounters.  As
relativistic effects are expected to affect the dynamics of the debris
in these types of encounters, a pertinent question is {\it can they be
  detected in observations?}

In the previous section we quantify the importance of relativistic
effects by calculating the ``lead time" of the relativistic curves
before the peak and the time delay of the peak relative to the
Newtonian ones. Specifically, we measure the delay in the peak of the
return rate of about 6 and 23 days for encounters MS2 and MS3,
respectively (Table~\ref{tab:return_rates}). In comparison, the
typical cadence of images and photometry taken as a part of the
Pan-STARRS1 Medium-Deep Field survey is every 3 nights
\cite{kaiser10}. Thus, these effects are in principle measurable with
the current transient sky surveys if the ballistic return rate of the
debris is proportional to the event light curve. This is an intriguing
possibility, especially for the MS2 class of events, which have been
considered numerous times in the literature as conventional Newtonian
encounters.  The relativistic effects in the return rate curve for WD
encounters are even more severe, but may be difficult to measure with
current surveys given the short time scale on which these events play
out ($< 1$~ day).

One important factor that prevents a simple interpretation of the
return rates of the debris in terms of the observed light curves is
the super-Eddington nature of the return rates. Namely, if simulated
return rates (which determine the supply rate of the gas) are translated
into the accretion rates onto the BH, they correspond to
super-Eddington accretion rates. If these in turn power
super-Eddington luminosities even for a brief period of time,
radiative feedback is likely to strongly affect the dynamics of the
accreting debris. That is, once the luminosity of the central source exceeds
the Eddington limit, the portion of the light curve above this
threshold (Fig.~\ref{figure:return_rates}) will be shaped by the
response of the debris to radiation pressure. As our simulations do
not account for interactions of matter and radiation, we cannot
determine whether signatures of relativistic effects will be preserved
above this threshold.

We note that relativistic features in the early, sub-Eddington phase
of evolution may still be preserved.  In the early times of the rate,
the relativistic curves have shallower slopes than the Newtonian
counterparts.  It is an interesting situation if an observed light
curve from a relativistic encounter was unknowingly modeled by a light
curve calculated from Newtonian simulations. In this case, the peaks
of the two curves may be arbitrarily shifted along the time axis until
they overlap but the slopes in the portion of the curves leading to
the peak would be discrepant. We indeed find an indication of this
behavior in the light curve of a tidal disruption event PS1-10jh for
which the early rise in the light curve has been observed in the
optical band with Pan-STARRS1 (see Fig.~2 in \cite{geza2012}). The
exact nature of this tidal disruption event has been disputed and two
different explanations have been offered in the literature. One group
of authors, including the authors of this paper, suggests that this is
a tidal disruption of a helium WD by a $\sim10^6\,M_{\odot}$ BH
\cite{geza2012,bogdanovic14}, while the other explains it in terms of
the disruption of a main sequence star by a $\sim10^7\,M_{\odot}$ BH
\cite{guil14}. Both are in agreement that the encounter is likely to
be relativistic. The case of PS1-10jh is interesting but unlikely to
be constrained by further observations because the unique spectral and
photometric features that can be used to infer its nature have faded
out of sight. Thus, the only remaining prospect to understand this
object may be through careful modeling of its relativistic light
curve, which we defer to future work.

Given a possibility that relativistic features in the light curve of
PS1-10jh have already been detected, it is of interest to consider
what kind of biases can be introduced to interpretation of tidal
disruption events if parametric fits based on Newtonian simulations
are used to model light curves from relativistic encounters.  We
address this issue by applying a simple model of the return rate
curves from our relativistic simulations by Newtonian fits by
\cite{guil2013}. We line up the relativistic and Newtonian curves in
such way that their peaks overlap, as it would be the case with a
majority of routines used to fit the observed data. Assuming a known
structure of the disrupted star, the timing and magnitude of the peak
are directly determined by the mass of the BH and strength of the
encounter. Solving for these properties in practice allows an
inference of the BH mass from tidal disruption events. Therefore, our
simple modeling procedure provides estimates for the BH masses which
we show in the last column of Table~\ref{tab:return_rates} as $M_{\rm
  est}$. Except for the MS1 run which does not exhibit significant
relativistic effects, we find that a simple model of relativistic
encounters with Newtonian parametric fit of the peak time leads to an
overestimate of the BH mass by tens of percent in the case of
disruptions of main sequence stars and a factor of few for WD
disruptions.  We nevertheless caution that more robust estimates of
the BH mass require the development of proper data analysis tools, a
worthwhile task that is out of the scope of this paper.

Given the increasing quality and detailed observational coverage of
tidal disruption events a question of how debris circularizes to form
an accretion disk becomes ever more pressing. In this work we show
that evolution in angular momentum is as important as the evolution in
orbital energy of the debris and that the first step towards
reconstruction of the debris orbits can be made only if both are
known. An important consequence of this is shown in
Fig.~\ref{figure:BH_frame_WD5} where both bound and unbound fluid
elements mix as they approach their periastrons. As this likely leads
to orbit crossing and collisions, it has important implications for
the evolution of debris disks.  While we obtain the orbital map of the
debris in the frame of the BH by propagating fluid elements
semi-analytically from the final snapshot of the simulation forward in
time, this approach does not account for the effects of
(magneto)hydrodynamics, self-gravity, or radiation transport. It thus
does not provide a final answer about how debris disks evolve but it
makes a point that angular momentum distribution as well as
relativistic effects are likely to play an important role.

\section{Conclusions}
\label{sec:conclusions}

In this study, we used a suite of Newtonian and relativistic
simulations of tidal disruption encounters of MS and WD stars with BHs
to investigate relativistic effects in the dynamics of debris. We
developed a local to BH frame transformation in order to calculate the
orbital parameters of the debris and used these to infer the return
rate of the debris as a function of time.  We evaluate the
relativistic effects in the orbital energy and angular momentum of the
debris as a function of BH mass and stellar type.  Severe relativistic
effects lead to the mixing and collision of fluid elements that are
both bound and unbound to the BH underlining the need for a full
hydrodynamic treatment to accurately capture the evolution of debris
accretion disks.  The two most pronounced signatures of relativistic
effects in the return rate of the debris to periastron are the gradual
rise and offset in the peak of the curve relative to the Newtonian
predictions. These are significant enough in the encounters of MS
stars and BHs that they can in principle be measured by the current
synoptic sky surveys, assuming that the return rate is proportional to
the event light curve.  Furthermore, with this assumption, if the
tidal disruption light curve from a relativistic encounter is simply
modeled with a Newtonian parametric fit of the peak time, this can
lead to an overestimate in the BH mass by a factor of $\sim {\rm
  few}\times0.1$ and $\sim {\rm few}$ for disruptions of the MS stars
and WDs, respectively.


\acknowledgments

R.M.C.~thanks the anonymous referees for providing valuable queries
for clarity.  The authors are grateful to the Kavli Institute for
Theoretical Physics for hosting the program, {\it A Universe of Black
  Holes}, where a portion of this work was completed. This research
was supported in part by the National Science Foundation under Grant
No. NSF PHY-1125915 and NSF AST-1333360. T.B. acknowledges the support
from the Alfred P. Sloan Foundation under Grant No. BR2013-016.

\appendix

\section{Expansion coefficients in FNC to BH frame transformation}
\label{sec:expansion}
The expansion coefficients $\sigma^{\ \ \mu}_{ij}$ and
$\kappa^{\ \ \ \mu}_{ijk}$ in
Eq.~\eqref{eqn:FNC_to_BH_coordinate_transformation} and
Eq.~\eqref{eqn:FNC_to_BH_velocity_transformation} are obtained by the
following procedure \citep{klei2008}.  The FNC frame center at
$X^{\mu}_{(0)}$ is located on a timelike geodesic $\mathcal{G}$
parametrized by proper time $\tau$ in the BH frame.  The relation
between the Fermi normal coordinates and the tetrad on $\mathcal{G}$ may be
expressed as the evaluation at $s=1$ of the solution $X^\mu(s)$ of the
initial value problem for a geodesic,
\ba\label{eqn:initial_value_problem_for_geodesic}
\nn && \frac{d^2 X^\mu}{ds^2}  +\Gamma^\mu_{\ \alpha\beta} \frac{dX^\alpha}{ds}\frac{dX^\beta}{ds} 
= 0, \\
&& X^\mu(0) = X^\mu_{(0)}, \qquad \frac{dX^\mu}{ds}(0) = x^i \lambda^{\ \mu}_i(\tau),
\ea
for spatial components $i=1,2,3$.  From the initial conditions, we express the
expansion for $X^\mu(s)$ as
\ba\label{eqn:expansion_for_point_in_BH_frame}
X^\mu (s) 
 & = & X^\mu_{(0)} + s a_1^\mu  + \tfrac{1}{2} s^2 a_2^\mu
 + \tfrac{1}{6} s^3 a_3^\mu + \cdots
\ea
where $a_1^\mu=\lambda^{\ \mu}_i x^i$ and $a_2$, $a_3$, $a_4$, and
$a_5$ are to be determined, and the connection coefficients as
\ba\label{eqn:expansion_for_connection_in_BH_frame}
\Gamma^\mu_{\ \alpha\beta}(X^\mu(s)) 
 & = & \Gamma^\mu_{\ \alpha\beta} \Big |_0 + \Gamma^\mu_{\ \alpha\beta,\gamma} \Big |_0 (X^\gamma(s) - X^\gamma_0) \\
\nn + \tfrac{1}{2}\Gamma^\mu_{\ \alpha\beta,\gamma\delta} & \Big |_0 & (X^\gamma(s) - X^\gamma_0) (X^\delta(s) - X^\delta_0) + \cdots,
\ea
with evaluations at $X^\mu_{(0)}$.  Substituting
Eq.~\eqref{eqn:expansion_for_point_in_BH_frame} and
Eq.~\eqref{eqn:expansion_for_connection_in_BH_frame} into the geodesic
equation \eqref{eqn:initial_value_problem_for_geodesic} and comparing
terms at each power of $s$, we have
\ba
\sigma^{\ \ \mu}_{ij} 
\nn & = & -\Gamma_{\ \alpha\beta}^{\mu}\Big |_0 \lambda_i^{\ \alpha} \lambda_j^{\ \beta}\\
\kappa_{ijk}^{\ \ \ \mu}
& = & \left ( 2 \Gamma_{\ \alpha\beta}^\nu \Gamma_{\ \nu\gamma}^\mu - \Gamma_{\ \alpha\beta,\gamma}^\mu \right ) \Big |_0 \lambda_i^{\ \alpha} \lambda_j^{\ \beta} \lambda_k^{\ \gamma}. 
\ea 

\section{Orbital parameters of the debris}
\label{sec:orb_param}

\subsection{Newtonian gravitational potential}

We consider the dynamics of the stellar debris in a Newtonian
gravitational potential described by time $t$ and spatial coordinates
$X^i=\{r,\theta,\phi\}$ and velocities $V^i=dX^i/dt=\dot{X}^i$.  For a
general orbit in a Newtonian gravitational potential $\Phi_N = -M/r$
(in geometrized units), an equation relating the specific orbital
energy $\epsilon$ and angular momentum $\ell$ is given by
\be\label{eqn:Newtonian_energy_relation} 
\epsilon = -\frac{M}{r} +
\frac{1}{2}\left ( \dot{r}^2 + \frac{\ell^2 + \mathcal{Q}}{r^2} \right
),
\ee
where $\ell = r^2 \sin^2\theta\dot\phi$ and $\mathcal{Q} = \ell^2 \cot^2\theta +
r^4 \dot{\theta}^2$ \cite{land1969}.  We parametrize the orbit in
terms of the semilatus rectum $p$ and eccentricity $e$ with
definitions for the turning points at periastron $R_p$ and apastron $R_a$,
\be\label{eqn:orbit_turning_points}
R_p = \frac{pM}{1+e}, \qquad \qquad R_a = \frac{pM}{1-e}.
\ee
Substituting Eq.~\ref{eqn:orbit_turning_points} into
Eq.~\ref{eqn:Newtonian_energy_relation} at $\dot{r}=0$, we relate the
specific orbital energy and angular momentum with the semilatus rectum
and eccentricity,
\be\label{eqn:Newtonian_specific_orbital_parameters_e_l}
|\epsilon| = \frac{M (1 - e^2)}{2p}, \qquad \ell = \sqrt{pM-\mathcal{Q}}
\ee
or
\be\label{eqn:Newtonian_specific_orbital_parameters_p_e}
p M = (\ell^2 + Q), \qquad e M = \sqrt{M^2 + 2 \epsilon(\ell^2+ \mathcal{Q})}.
\ee
Note that the semi-major axis $a$ is related to these quantities by $a
= (R_p+R_a)/2 = p/(1-e^2) = M/(2|\epsilon|)$.  

The time evolution along the orbit is obtained by re-writing
Eq.~\ref{eqn:Newtonian_energy_relation} as
\be\label{eqn:Newtonian_time_evolution_1} 
\frac{dt}{dr} = \sqrt\frac{a}{M} \frac{r}{\sqrt{a^2 e^2 - (r-a)^2}}
\ee
We introduce a radial phase angle $\xi = [0,2\pi]$ to parametrize the orbit \cite{land1969},
\be\label{eqn:Newtonian_radial_phase}
r = a(1-e\cos\xi).
\ee
Substituting Eq.~\eqref{eqn:Newtonian_radial_phase} into
Eq.~\eqref{eqn:Newtonian_time_evolution_1}, we have a regularized equation for the time evolution,
\be\label{eqn:Newtonian_time_evolution_2}
\frac{dt}{d\xi} = \sqrt{\frac{a^3}{M}} (1-e\cos\xi).
\ee
We regularize the first-order equation of motion in $\theta$ (from
$\mathcal{Q}$ in Eq.~\eqref{eqn:Newtonian_energy_relation}) in a
similar manner to \cite{hugh2001}, by defining
$\mathcal{Z}=\cos^2\theta$ and introducing a phase $\varphi=[0,2\pi]$
such that $\mathcal{Z} = \mathcal{Z}_0 \cos^2\varphi$.  Turning points
in $\theta$ occur at $\mathcal{Z}_0 = \mathcal{Q}/(\mathcal{Q} +
\ell^2)$, where $\theta_\pm = \cos^{-1}(\pm \sqrt{\mathcal{Z}_0})$.
We then have a first-order equation for $\varphi$,
\be\label{eqn:polar_phase}
\frac{d\varphi}{dt} = \frac{1}{r^2}\sqrt{\frac
{\mathcal{Q} - \mathcal{Z} (\mathcal{Q} + \ell^2)}
{\mathcal{Z}_0-\mathcal{Z}}
},
\ee
which parametrizes motion in $\theta =
\cos^{-1}(\cos\varphi/\sqrt{\mathcal{Z}_0})$.  Motion in $\phi$ is
obtained in a straightforward manner by integrating the first-order
equation from $\ell$ in Eq.~\eqref{eqn:Newtonian_energy_relation}.

\subsection{Schwarzschild spacetime}

We consider the dynamics of the stellar debris in the spacetime of a
Schwarzschild BH described by coordinates $X^\mu=(t,r,\theta,\phi)$,
parametrized by proper time $\tau$, and four-velocity
$U^\mu=dX^\mu/d\tau = \dot{X}^\mu$.  A relation for general
orbits on Schwarzschild is given by
\be\label{eqn:Schwarzschild_specific_orbital_energy} 
\epsilon^2 =
\dot{r}^2 + f \left ( 1 + \frac{\ell^2+\mathcal{Q}}{r^2}
\right), 
\ee 
in terms of the specific orbital energy $\epsilon = f \dot{t}$,
specific angular momentum $\ell = r^2 \sin^2\theta \dot{\phi}$, and
Carter constant $\mathcal{Q}= \ell^2 \cot^2\theta + r^4
\dot{\theta}^2$, where $f=1-2M/r$ \cite{misn1973}.  

Substituting Eq.~\ref{eqn:orbit_turning_points} into
\eqref{eqn:Schwarzschild_specific_orbital_energy}, we relate the
specific orbital energy and angular momentum with the semilatus rectum
and eccentricity,
\ba\label{eq:relativistic_specific_orbital_parameters}
\epsilon^2 
\nn &=& \frac{(p-2-2e)(p-2+2e)}{p(p-3-e^2)} , \\
\ell^2 &=& \frac{p^2M^2}{p-3-e^2} - \mathcal{Q}.
\ea
Instead of inverting
\eqref{eq:relativistic_specific_orbital_parameters} to obtain
relations for $\{e,p\}$ in terms of given quantities
$\{\epsilon,\ell\}$, we make use of the third root of the effective
potential \cite{cutl1994},
\be\label{eq:third_root}
r_3 = \frac{2 M p}{p-4}.
\ee
Given $\{\epsilon,\ell\}$, we numerically solve for this root with a
Newton-Raphson method \cite{pres1986} within the interval
$\{1.0M,4.0M\}$ to obtain $p$ from Eq.~\eqref{eq:third_root}.  Using
Eq.~\eqref{eq:relativistic_specific_orbital_parameters}, we obtain $e$
and then $R_p$ and $R_a$ from Eq.~\eqref{eqn:orbit_turning_points}.

The orbital radius of the geodesic is parametrized by the radial phase
angle $\chi$,
\be\label{eqn:parametrization_radius} 
r(\chi) =
\frac{pM}{1+e\cos\chi}.  
\ee
The evolution along the orbit in Schwarzschild time is given by \cite{cutl1994}
\ba\label{eqn:darwin_parametrization_time}
\frac{dt}{d \chi} 
&=& \frac{p^2 M}{(p-2-2 e \cos\chi)(1 + e \cos\chi)^2} \nn \\
& & \quad \times 
\left[ \frac{(p-2)^2 - 4 e^2}{p - 6 - 2 e \cos \chi} \right]^{1/2}.
\ea
Motion in $\theta$ and $\phi$ is obtained similar to the Newtonian
case.


\bibliographystyle{apsrev4-1}
\bibliography{FNC_debris}

\end{document}